\documentclass{aa}
\usepackage{natbib}
\usepackage{graphicx}
\usepackage{txfonts}
\usepackage{hyperref}
\usepackage{url}
\usepackage{xcolor}
\hypersetup{
colorlinks,
linkcolor={red!80!black},
citecolor={blue!80!black},
urlcolor={blue!80!black}
}

\def\cm3{cm$^{-3}$}
\def\kms{km~s$^{-1}$}

\def\msun{M$_{\odot}$}

\def\one{\ts {\,\sc i}}
\def\two{\ts {\,\sc ii}}

\def\beq{\begin{equation}}
\def\eeq{\end{equation}}

\def\lesssim{\mathrel{\hbox{\rlap{\hbox{\lower4pt\hbox{$\sim$}}}\hbox{$<$}}}}
\def\gtrsim{\mathrel{\hbox{\rlap{\hbox{\lower4pt\hbox{$\sim$}}}\hbox{$>$}}}}

\def\isoni{$^{56}{\rm Ni}$}

\def\one{{\,\sc i}}
\def\two{{\,\sc ii}}

\def\v1d{{\sc v1d}}

\def\mesa{{\sc mesa}}
\def\cmfgen{{\sc cmfgen}}

\newcommand{\iso}[2]{\ensuremath{^{#1}\rm{#2}}}

\def\aj{AJ}

\def\apj{ApJ}
\def\apjs{ApJS}
\def\apjl{ApJL}
\def\aap{A\&A}
\def\araa{ARA\&A}

\def\mnras{MNRAS}
\def\nat{Nature}

\def\nifs{\iso{56}Ni}

\begin{document}

   \title{Impact of clumping on core-collapse supernova radiation}
   \titlerunning{Clumping and core-collapse SN radiation}

\author{Luc Dessart\inst{\ref{inst1}}
  \and
   D. John Hillier\inst{\ref{inst2}}
   \and
   Kevin D. Wilk\inst{\ref{inst2}}
  }

\institute{
Unidad Mixta Internacional Franco-Chilena de Astronom\'ia (CNRS, UMI 3386),
    Departamento de Astronom\'ia, Universidad de Chile,
    Camino El Observatorio 1515, Las Condes, Santiago, Chile\label{inst1}
\and
    Department of Physics and Astronomy \& Pittsburgh Particle Physics,
    Astrophysics, and Cosmology Center (PITT PACC),  University of Pittsburgh,
    3941 O'Hara Street, Pittsburgh, PA 15260, USA.\label{inst2}
  }

   \date{Received; accepted}

  \abstract{
  There is both observational and theoretical evidence that the ejecta of core-collapse
  supernovae (SNe) are structured. Rather than being smooth and homogeneous, the material is
  made of over-dense and under-dense regions of distinct composition.
  Here, we explore the effect of clumping on the SN radiation during the photospheric phase
  using 1-D non-local thermodynamic equilibrium radiative transfer and an ejecta model arising
  from a blue-supergiant explosion (yielding a Type II-peculiar SN). Neglecting chemical
  segregation, we adopt a velocity-dependent volume-filling factor approach that
  assumes that the clumps are small but does not change
  the column density along any sightline. We find that clumping boosts the recombination
  rate in the photospheric layers, leading to a faster recession of the photosphere, an increase
  in bolometric luminosity, and a reddening of the SN colors through enhanced blanketing.
  The SN bolometric light curve peaks earlier and transitions faster to the nebular phase.
  On the rise to maximum, the strongest luminosity contrast between our clumped
  and smooth models is obtained at the epoch when the photosphere has receded to ejecta
  layers where the clumping factor is only 0.5 -- this clumping factor may be larger in Nature.
  Clumping is seen to have a similar influence in a Type II-Plateau SN model.
  As we neglect both porosity and chemical segregation our models
  underestimate  the true impact of clumping.
  These results warrant further study of the influence of clumping on the observables
  of other SN types during the photospheric phase.
  }

\keywords{
  radiative transfer --
  radiation: dynamics --
  supernovae: general --
  supernova: individual: SN\,1987A
}
   \maketitle

\section{Introduction}

The diversity of core collapse supernova (SN) light curves arises from variations in the properties of the shocked progenitor envelope,
and the subsequent balance between heating (radioactive decay
and recombination) and cooling  (expansion and radiation)
processes (e.g. \citealt{FA77}).
In the explosion of red-supergiant (RSG) stars which lead to Type II-Plateau SNe, the escaping radiation
is dominated for several months by the release of the original shock-deposited energy,
which is modestly degraded by expansion cooling.
In the explosion of the more compact blue-supergiant (BSG) stars, which lead to Type II-peculiar SNe,
the stronger expansion cooling produces a fainter SN at early times. The timescale and rate at which
the SN subsequently brightens bears critical information about the
mass and extent of the hydrogen envelope, the amount and distribution of \nifs,
ejecta symmetries, and other progenitor and explosion properties
(e.g. \citealt{sn1987A_rev_90}; \citealt{wongwathanarat_15_3d}).
Because of the massive and more extended H-rich envelope this information is harder to decipher
in Type II-Plateau SNe, although some information can be gleaned from polarization studies
(e.g. \citealt{leonard_04dj_06}).

There is both observational and theoretical evidence that the ejecta of
core-collapse SNe, and SN\,1987A in particular, is structured on both large and small
scales.  Current radio observations
of CO and SiO molecular line emission from the innermost regions of SN\,1987A
reveal a clumpy asymmetric structure \citep{abellan_87A_17}.
Nebular phase spectra require significant macroscopic mixing, often combined with a clumpy
structure \citep{fransson_chevalier_89, spyromilio_87a_90,li_87A_93,jerkstand_87a_11,jerkstrand_04et_12}.
A clumpy structure is also suggested by the observed fine-structure in the H$\alpha$ line profile
at early times \citep{hanuschik_87a_88}. This `Bochum' event is further supported by
the direction-dependent spectra of SN\,1987A observed via light echoes \citep{sinnott_87a_13}.
Integral field spectroscopy applied to a selection of near-infrared lines also suggests
a large scale asymmetry of the SN\,1987A ejecta \citep{kjaer_87a_10}.
The smooth rising optical brightness and the high-energy radiation observed after about 200\,d
in SN\,1987A suggest the mixing of \nifs\ out to $3000-4000$\,\kms\
(e.g. \citealt{sn1987A_rev_90}).
Numerical simulations of core-collapse SN explosions suggest a strong breaking of spherical
symmetry on small and large scales, both from the intrinsic multi-dimensional nature of the
explosion mechanism and the shock wave propagation in a stratified massive star progenitor
\citep{muller_87A_91,kifonidis_00,sasi_03,wongwathanarat_15_3d}.

Despite the widespread knowledge that core-collapse SN ejecta are clumpy, light curve
and spectral calculations during the photospheric phase generally assume a smooth ejecta.
It is therefore of interest to investigate the impact of clumping on SN observables
during the photospheric phase, particularly since multiple effects may produce similar
changes in the light curve and spectra.

In the next section, we present our numerical approach.
In Section~\ref{sect_cl}, we discuss the treatment of clumping in our radiative transfer calculations,
the relevant clumping scales for continuum and line radiation, as well as the limitations
of our approach.
In Section~\ref{sect_res}, we compare the smooth and clumped models Bsm and Bcl,
including the differences in bolometric light curve, color, and spectral evolution.
A comparison of the models with observation is made in Section~\ref{sect_comp_obs}.
In this section we also argue that the early rise in SN\,1987A may be in part
driven by ejecta clumping, rather than \nifs\ mixing alone as generally stated.
Clumping can thus modulate the rate of energy release from Type II SN ejecta,
impacting the photospheric phase duration and brightness.
In Section~\ref{sect_disc}, we summarize our results and discuss how clumping might
alter current inferences of ejecta properties across SN types and how this alteration might
impact our model of the progenitor stars.

\section{Numerical approach}
\label{sect_setup}

 In this work, we discuss our results for a BSG progenitor.
  This model corresponds to a zero-age main sequence star
  of 15\,\msun\ evolved with \mesa\ \citep{mesa3}
  at a metallicity of 10$^{-7}$.  With this very low metallicity the massive star
  model reaches core collapse as a BSG star
  without the need for any tinkering of convection parameters etc.
  We simulate the explosion with \v1d\ \citep{livne_93}.
  The explosion is triggered by depositing a total
  of $1.6 \times 10^{51}$\,erg at a constant rate for 0.5\,s and
  over a region of 0.05\,\msun\ above a lagrangian mass of 1.55\,\msun.
  We apply a strong chemical mixing of all elements. This mixing
  brings core metals to the metal-poor outer ejecta, and thus largely erases the
  original low metallicity of the progenitor star (Fig.~\ref{fig_prog_comp}).
  At 1\,d, this model is remapped into \cmfgen\ \citep{HD12} and followed
  until the onset of the nebular phase using the standard procedure \citep{d13_sn2p}.
  Metals like Ba, not included in \mesa\ nor in \v1d, are given a constant mass fraction equal
  to the corresponding LMC metallicity value.
  Metals between Ne and Ni that are included in \mesa\ and \v1d are given a minimum
  mass fraction equal to the corresponding LMC metallicity value (this applies to the most
  abundant stable isotope for each element).
  The LMC value is adopted here since it is the metallicity relevant for SN\,1987A,
  to which our model will be compared. Furthermore, SN\,1987A
  is the prototypical Type II-peculiar SN, and such BSG explosions seem to occur
  at LMC-like metallicities \citep{taddia_2pec_z_13}.
   This initial \cmfgen\ model has a mass of 13.22\,\msun,
  a kinetic energy of $1.26 \times 10^{51}$\,erg, and an initial \nifs\ mass of 0.084\,\msun.

 \section{Clumping}
\label{sect_cl}

\subsection{Numerical treatment of clumping}

  To treat clumping we assume that the ejecta are composed of clumps that occupy a
  fraction $f_{\rm vol}$ of the total volume. The clumps are assumed to be `small'
  (see Sect.~\ref{Sect_consider_clumps}), there is no interclump medium,
  and the velocity remains unchanged (i.e., clump and inter-clump media move
  at the same speed at a given radius and post-explosion time).
  With these assumptions the  model density is
  simply scaled by a factor of $1/f_{\rm vol}$, while opacities and emissivities  (which
  are computed with populations and temperature deduced for the clumps) are all scaled by a factor
  of $f_{\rm vol}$. These assumptions leave column densities unchanged but have
  a direct influence on processes that depend on the density squared (e.g., free-free),
  and an indirect influence on the radiation field because of the sensitivity of
  the kinetic equations to density.

The volume-filling factor approach adopted has been implemented in \cmfgen\ \citep{1996LIACo..33..509H,hm99}
for  two decades,  is also implemented in other transfer codes such as {\sc fastwind}
\citep{2006A&A...454..625P} and {\sc PoWR} \citep{1998A&A...335.1003H}, and
has been widely used in the studies of massive stars and their stellar winds.
There is strong theoretical evidence that the winds of hot stars are clumped due
to instabilities arising from radiation driving \citep{owocki_rybicki_84,ocr88}.  Observations, of both O and W-R stars,
also indicate the clumping is important \citep{eversberg_zpup_88,moffat_88,hillier_91}.

The volume-filling factor approach is somewhat similar to assuming a radial compression of material into dense spherical shells, occupying a fraction $f_{\rm vol}$ of the total volume.
However, due to both radiative transfer effects (which would allow for optical depth effects)
and the influence of density variations, such shells would show strong variations in populations across the shells. Shell-like models also incorporate an interclump medium. Shell-like models have been explored for O and  W-R stars \cite{1991A&A...247..455H,2008cihw.conf...93H}.

   Multi-dimensional simulations of core-collapse SN explosions suggest a greater
   level of clumping as one progresses deeper in a core-collapse SN ejecta
  \citep{muller_87A_91,wongwathanarat_15_3d}.
  Hence, the adopted clumping factor is a function of the ejecta velocity $V$ and takes the form :
  \begin{equation}
     f_{\rm vol}(V) = 1 + (a-1) \exp(-X^2) \, \, ; \,\, X = (V-V_{\rm min}) / b  \, , \label{eq_cl}
 \end{equation}
  where $V_{\rm min}$ is the minimum ejecta velocity.
  With this choice, the inner ejecta is clumped ($f_{\rm vol}(V) = a$, with $a<$\,1) and becomes
  progressively smooth  outwards ($f_{\rm vol}(V \rightarrow \infty)=$\,1).
  The transition region from clumped to smooth ejecta is controlled by $b$.
  The choice for $a$ and $b$ is somewhat arbitrary. \citet{muller_87A_91}
  report density contrasts close to 10 in their simulations (i.e. $a\sim$\,0.1).
  For $b$, we chose the ejecta velocity that corresponds roughly to what used
  to be the edge of the He core in the progenitor star.
  Using our BSG progenitor model, we study two variants of the same explosion
  model. Model Bsm has a smooth ejecta ($a=$\,1)  and model Bcl has a clumped
  ejecta (with $a=$\,0.1 and $b=$\,4000\,\kms).
We assume that clumping does not evolve with time.

    To start the clumped models,
  we chose a post-explosion time sufficiently early that the photosphere
  in the smooth model is at a velocity where $f_{\rm vol}$ is close to unity
  in the clumped model.
  The emerging SN radiation in the clumped model is therefore not affected
  by clumping initially. We then stepped in time in the usual manner and
  with the same model parameters as in the smooth model.
  Throughout this work, we assumed local deposition of radioactive decay energy.

  \begin{figure}
    \includegraphics[width=\hsize]{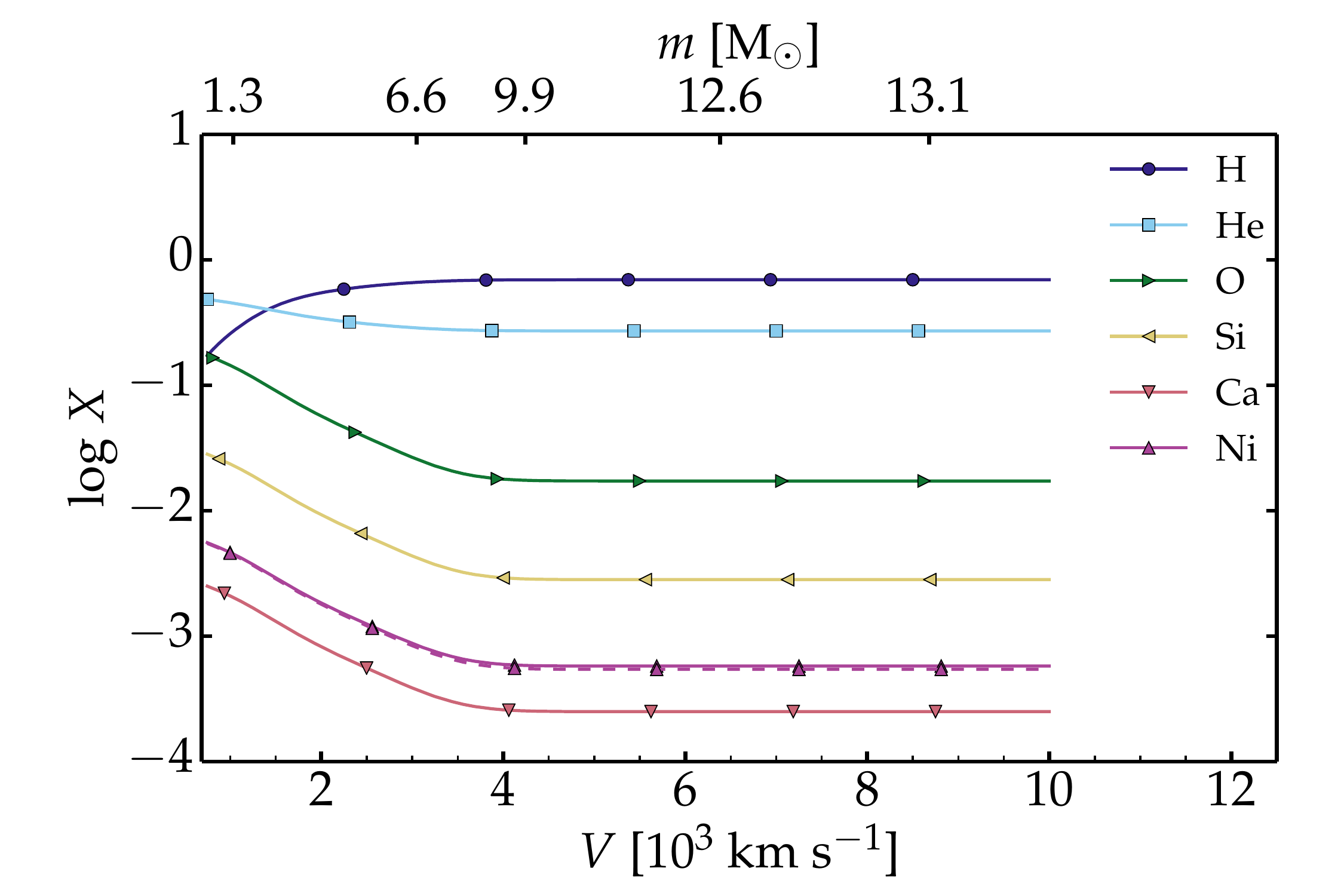}
    \caption{Chemical composition of the BSG explosion model used
      for the Bsm (smooth) and Bcl (clumped) simulations with \cmfgen\
      (Section~\ref{sect_res}).
      The $x$-axis is cut at 10000\,\kms\ -- the composition is uniform
      above that limit and up to the maximum ejecta velocity of 30000\,\kms. }
    \label{fig_prog_comp}
  \end{figure}

  \subsection{Considerations on clumping}
\label{Sect_consider_clumps}

Clumping will impact continuum and line photons differently. At large optical depths,
the change to the opacity caused by clumping is largely irrelevant since the diffusion
time is much larger than the age of the SN, and hence there is effectively
no diffusion of radiant energy  from regions of high optical depth. What matters most are
the photospheric layers in which diffusion occurs, and from which photons escape.

For continuum processes, our approach assumes that the clumps are `small' and optically thin,
or equivalently,  that the clumps are much smaller than a photon mean free path. In this case the relevant scale is $1/\chi$, where $\chi$ is the opacity. Assuming that the density at a radius $R$ varies as $R^{-n}$ the scale length at the photosphere is $\sim  R/(n-1)$. In our BSG explosion model, we have $n=9$ above 2000\,\kms. If we assume
the volume-filling factor is $f_{\rm vol}$, we require that the clumps are much smaller (say a factor of
10) than $R f_{\rm vol}/10$. Adopting $f_{\rm vol}=0.1$ and a photospheric radius  $\geq 10^{15}$cm
(typically valid, for example, between 10 and 160\,d) the clumps should have
a size  $\sim$\,10$^{12}$\,cm or smaller.

For line processes the relevant scale is the Sobolev length. The Sobolev length is defined as
  $V_{\rm th}/(dV/dR)$ $(=R  V_{\rm th}/V)$ where $V_{\rm th}$ is the intrinsic line thermal width, which varies from $\sim$\,1\,\kms\ for Fe to $\sim$\,10\,\kms\  for H in a Type II photosphere at the recombination epoch.  For $R=10^{15}$ cm,  $V_{\rm th}=1$\,\kms\ (i.e. an Fe line with no micro-turbulence),  and $V_{\rm phot}=5000$\,\kms, the Sobolev length is $2\times 10^{11}$ cm -- i.e., about the size of the Sun. The approach we use to incorporate clumping into \cmfgen\  effectively assumes that the clumps are much smaller than a Sobolev length.While this scale is smaller than the continuum requirement, it is still large. In practice, we use a turbulent velocity of 50\,\kms\ in the present \cmfgen\ simulations. This tends to enhance the Sobolev length and make it less species/ion dependent.

As noted earlier,  there is strong theoretical and observational evidence that the
winds of hot stars are clumped with the clumps having a size of a few Sobolev lengths. However there are arguments about the nature of the clumps and the importance of porosity. In stellar winds we can also have  porosity in
velocity space \citep{owocki_08} -- some directions may not probe all velocities
or some velocity regions may have a much lower density with little impact on photon escape.
The properties of the interclump medium also impact the morphology of P~Cygni profiles,
in particular whether they exhibit a `black absorption' (i.e., no residual flux in the
P-Cygni absorption trough; e.g., \citealt{2010A&A...510A..11S}). An approximate approach to solving the non-LTE transfer problem in such cases is outlined by \citet{2018arXiv180511010S}. Despite the limitations of the volume-filling factor approach, it has been used quite successfully in modeling both O \cite[e.g.,][]{2003ApJ...595.1182B,2012A&A...544A..67B} and W-R stars \cite[e.g.,][]{2015A&A...581A..21H}.

The clumping \cmfgen\ uses corresponds to an ejecta composed of numerous small-scale inhomogeneities, covering all ejecta-centered directions, rather than a few large scale structures. The properties of these small-scale clumps may be complex. For example, \isoni\  bubbles will cause regions of enhanced density, and they themselves will be in regions more rarified than what is assumed by a smooth hydrodynamical
model.

Our treatment of clumping is different from the statistical approach of \citet{jerkstand_87a_11}
in two important ways. First, our clumping does not alter the column density,
and hence the electron-scattering optical depth (ignoring changes in ionization) and
the escape of gamma-ray photons.  The second characteristic of clumping not considered
in \cmfgen\ is chemical segregation.  In core-collapse SNe, mixing is expected to be macroscopic,
leading to a shuffling of the pre-SN  `onion-shell' structure by advection on a large scale,
with little or no microscopic mixing. At a given radius (or velocity) in a macroscopically mixed model, one encounters material that used to be located at distinct radii (or velocities) in the original, unmixed, model, and
with its original composition.

%in this unmixed model.

The assumption of microscopic instead of macroscopic mixing can, in some cases, have
dramatic effects on nebular spectra. As our own modeling has shown, and was first
demonstrated for SN 1987A by \citet{fransson_neb_87A_87},
microscopic mixing of Ca into the O zone
can lead to both an enhancement in the strength of the [Ca\two] lines,
and a corresponding reduction in the strength of the [O\one] lines.
Because the [Ca\two] lines are a more efficient coolant
than the [O\one] lines, the amount of Ca in the O zone need not be very large
to cause the [O\one] line flux to weaken substantially.

  In \cmfgen,
  we assume both macroscopic and microscopic mixing. This means that each radial (or velocity)
  shell in a \cmfgen\ model (whether smooth or clumped) is fully mixed. While this seems a severe
  shortcoming of our approach, the \cmfgen\ results obtained at nebular times for
  Type II SNe \citep{d13_sn2p} are a close match to the observations \citep{silverman_neb_17},
  not obviously inferior in quality to simulations that treat chemical segregation and clumping
  \citep{jerkstrand_04et_12}.  This weak sensitivity implies that nebular-phase spectra cannot
  set a robust constraint on chemical segregation and clumping in the inner ejecta.

A more realistic treatment of clumping will require a 3-D approach, and should allow
for a variety of clump properties (abundances, filaments, broad protuberances, or blobs) on
different scales, and for the interclump medium (assumed to be a vacuum in \cmfgen).
In such a scenario porosity may be important, that is, `clumping' might lead
to a reduction of the optical depth along some sight lines.

\section{Results}
\label{sect_res}

In this section, we compare the results of a clumped model with those of an otherwise
identical but smooth model. We study the bolometric luminosity, optical colors, optical spectra,
and the ejecta properties and the photosphere as a function of time until the onset
of the nebular phase.

\begin{figure}
  \includegraphics[width=\hsize]{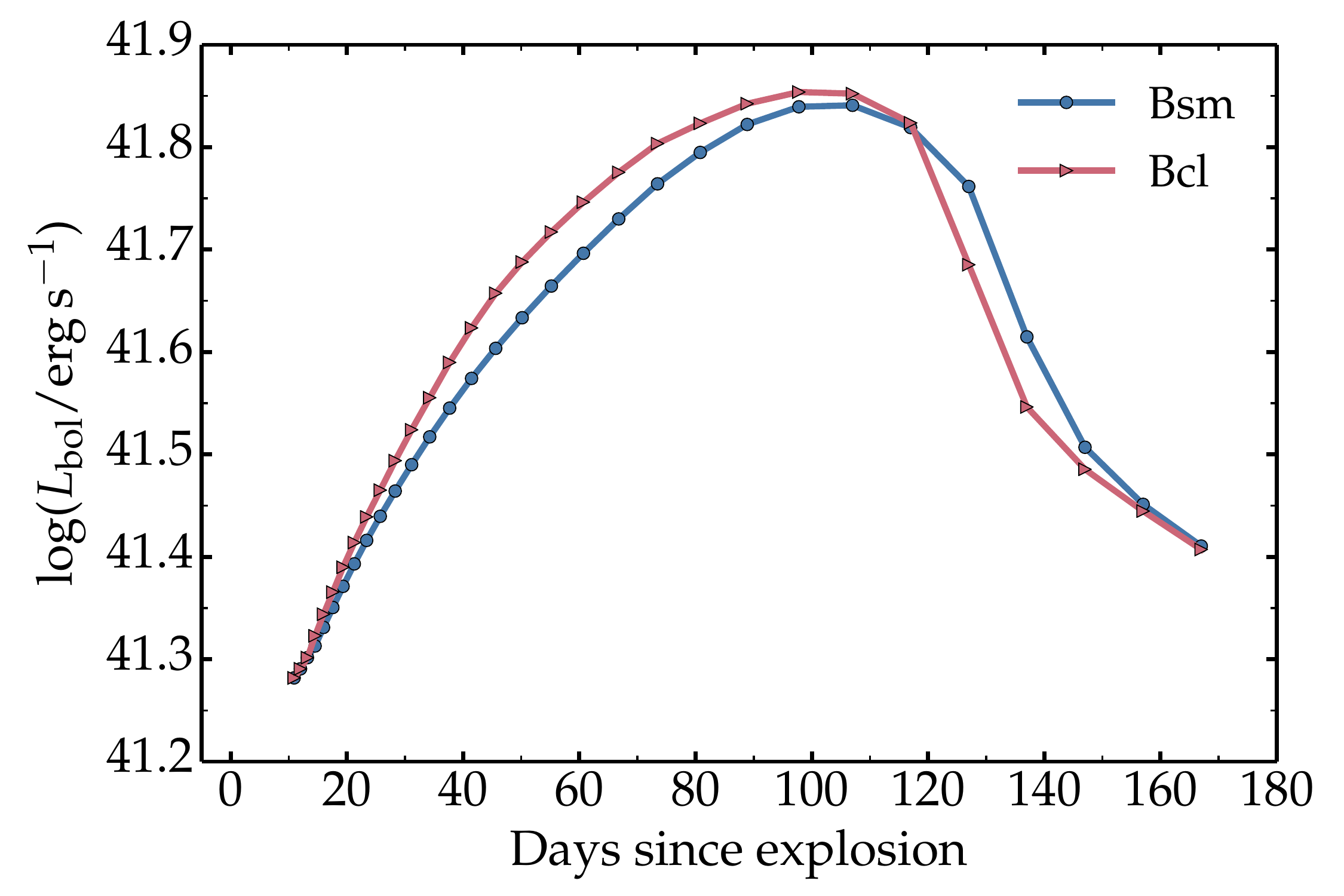}
  \includegraphics[width=\hsize]{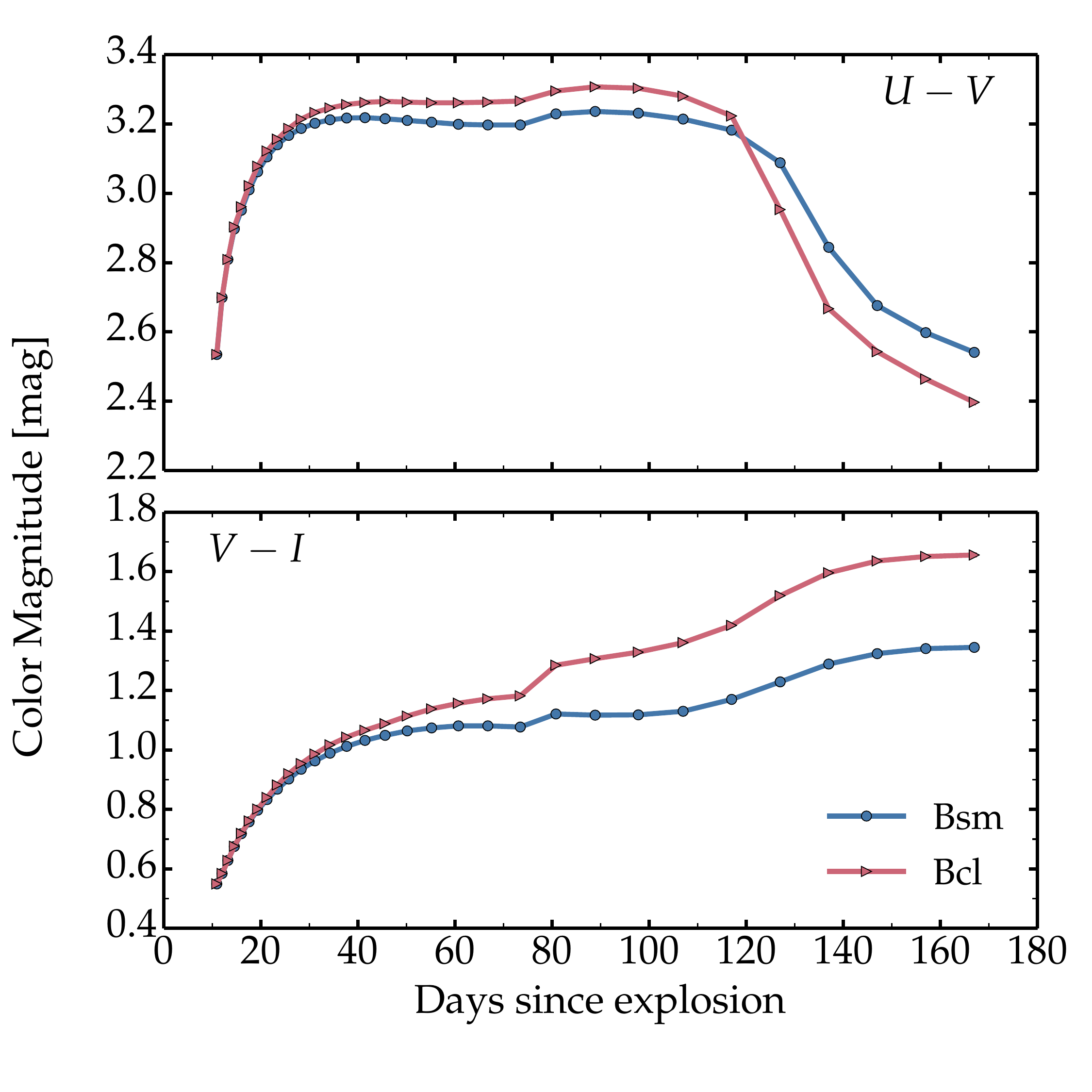}
  \caption{Bolometric light curve (top) and  color evolution (lower panels) for
    models Bsm and Bcl.
    \label{fig_photometry}
  }
\end{figure}

Figure~\ref{fig_photometry} shows the bolometric light curve (top) and
the $U-V$ and $V-I$ colors (bottom panels) from 10\,d to 170\,d after explosion
for the smooth and clumped models Bsm and Bcl.
The two models are initially indistinguishable but subsequently depart from
each other, the clumped model Bcl becoming more luminous and redder in the optical
than the smooth model Bsm.
The offset in bolometric luminosity grows to $10-15$\,\% at 50\,d before
decreasing again as the clumped model reaches its bolometric maximum earlier, by
about 10\,d. The excess radiation released at early times turns into a deficit of radiation
after maximum, causing the clumped model to fade earlier from maximum (with respect
to the time of explosion). The clumped model turns optically thin earlier.
At 170\,d, the two models are optically thin in the continuum (some lines may remain optically
thick for much longer) and have the same bolometric
luminosity.\footnote{Because we assume a local energy deposition, the nebular-phase
luminosity of both models is identical. But with our treatment of clumping, which preserves
the column densities, the $\gamma$-ray energy deposition (and hence the luminosity)
would be equal in both models even if we allowed for non-local energy deposition.}
Rather than impacting the time-integrated bolometric luminosity, clumping influences
the rate at which radiation is released from the SN ejecta.
Up to maximum, the clumped model is consistently redder in the optical than the smooth
model. This persists in $V-I$ after maximum, while the clumped model shows a bluer
$U-V$ color as the ejecta turns nebular.

\begin{figure}
  \vspace{-0.46cm}
  \includegraphics[width=\hsize]{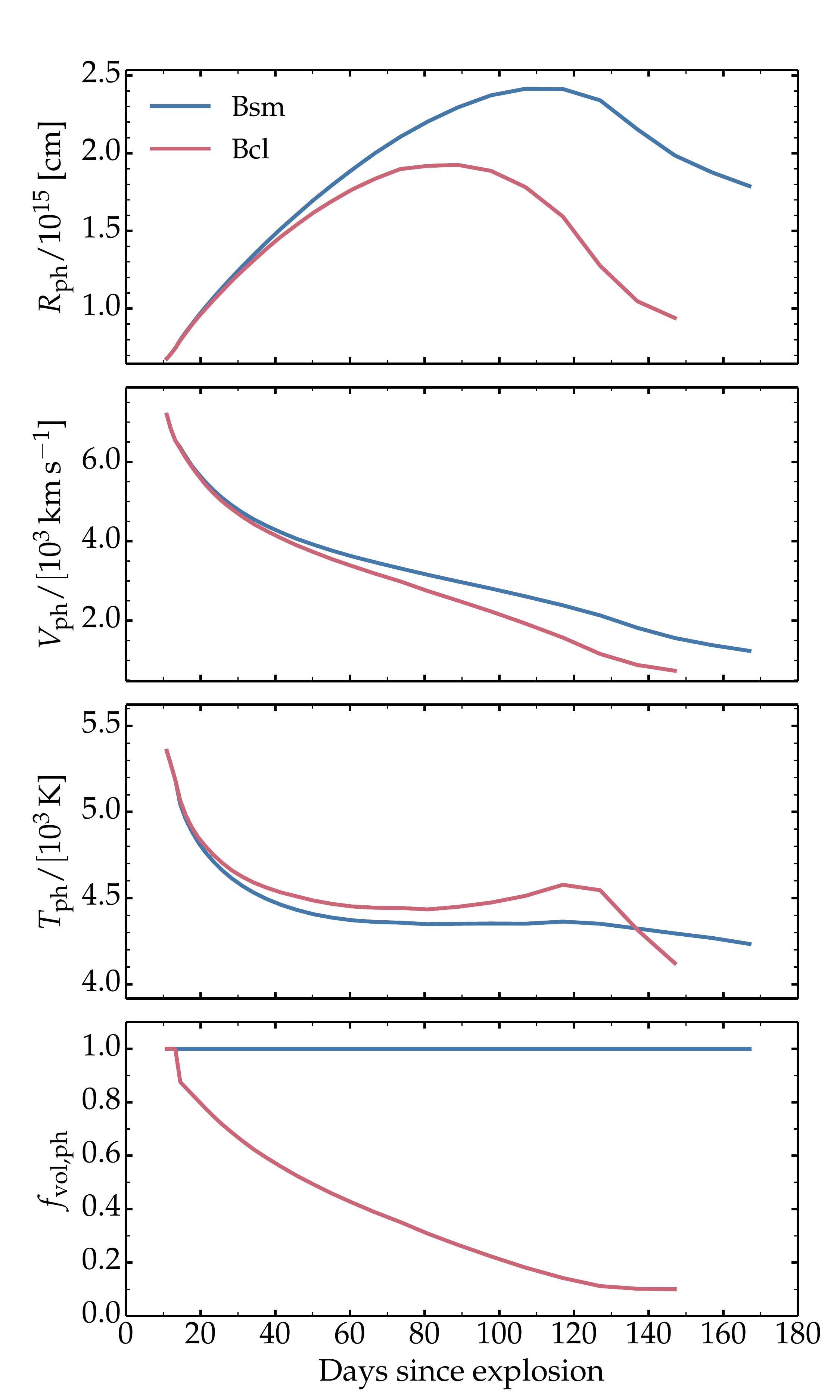}
  \caption{
    Evolution of the radius, the velocity, the temperature, and the clumping factor
    at the photosphere of models Bsm and Bcl.
    \label{fig_phot_prop}
  }
\end{figure}

Figure~\ref{fig_phot_prop} shows some photospheric properties for models Bsm and Bcl,
which illustrates the origin of the luminosity boost in the clumped model.
In both models, the photosphere recedes toward the inner ejecta layers as time passes
but in the clumped model this recession is faster. Consequently, the photospheric velocity
drops faster with clumping, yielding smaller photospheric radii.
The faster recession allows more trapped energy to be released, boosting the luminosity
on the way to bolometric maximum.
Given the smaller photospheric radii, this occurs through a greater photospheric temperature.
The greater offset in bolometric luminosity takes place when the clumping factor at
the photosphere is only 0.5 (over-density of a factor of 2 compared to the smooth model).
Numerical simulations (e.g., \citealt{muller_87A_91}) suggest much greater clumping values,
which suggests that the effect seen here could be greater in Nature.

\begin{figure}
  \vspace{-0.46cm}
  \includegraphics[width=\hsize]{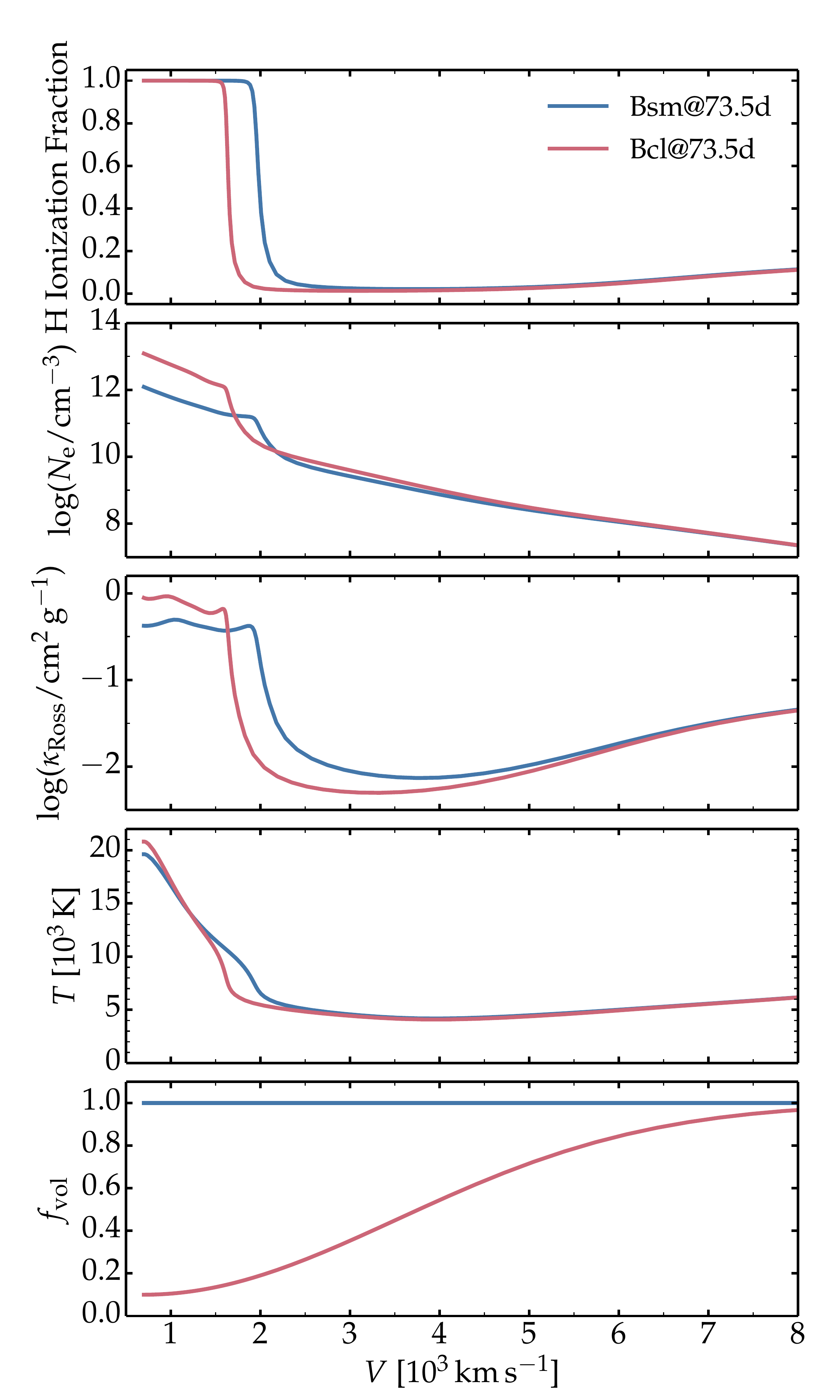}
  \caption{
    Comparison of ejecta properties for models Bsm (solid) and Bcl (dashed) at 73.5\,d after explosion.
    \label{fig_sm_cl_snap}
  }
\end{figure}

\begin{figure}
  \includegraphics[width=\hsize]{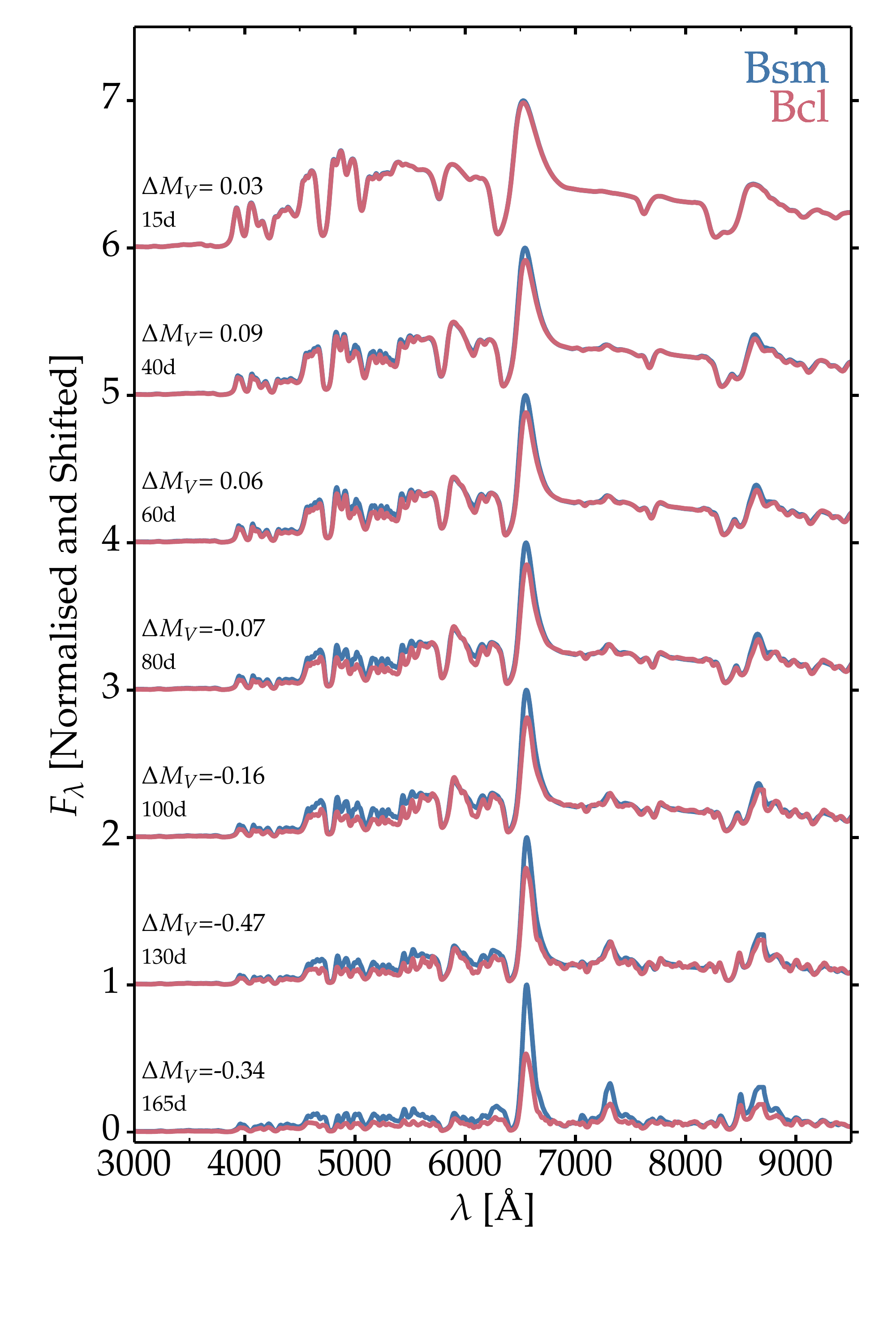}
  \vspace{-1.5cm}
  \caption{
    Comparison of the spectral evolution for  models Bsm and Bcl.
    At each epoch, the Bsm model is divided by the peak H$\alpha$ flux and shifted
    vertically by one unit, and the Bcl model is normalized to the Bsm model at 6800\,\AA.
    The labels at left indicate the post-explosion time and the $V$-band magnitude offset
    between the two models.
    \label{fig_spec_sm_cl}
  }
\end{figure}

Figure~\ref{fig_sm_cl_snap} shows some ejecta properties of the clumped and smooth models
Bcl and Bsm at 73.5\,d after explosion. The profiles look similar in both models, which exhibit a strong
ionization front. The jump in the H ionization fraction coincides with a jump in electron density,
in Rosseland-mean opacity and in temperature. Note that in both the clumped and the smooth models,
the photosphere is located well above this ionization front (by about 1000\,\kms\ at that time;
see Fig.~\ref{fig_phot_prop}). This feature arises from the ionization freeze-out in the outer ejecta,
allowing the electron-scattering optical depth to be $5-10$ at the ionization front.
However, in the clumped model,
the ionization front has receded $\sim$\,500\,\kms\ deeper in the ejecta, and will eventually
reach the base of the ejecta 10\,d earlier than in the smooth model Bsm.
Figure~\ref{fig_sm_cl_snap} also shows that the clumped model has a higher Rosseland-mean
opacity in optically thick regions, and a higher Rosseland-mean optical depth (1070
versus 620 in the smooth model) while the reverse is true for the electron-scattering
optical depth (201 versus 230 in the smooth model). This offset arises because density-squared
dependent opacity processes are boosted in the clumped model, while those that are linear
(like electron scattering) are not (the clumped ejecta has a lower electron-scattering
optical depth because its ionization is lower).
So, counter-intuitively, the clumped model, with its greater Rosseland-mean optical depth,
has a shorter photospheric phase. This arises because the total optical depth is not the
relevant one for diffusion. In Type II SN ejecta, diffusion occurs mostly through the photospheric
region.

Let $X^{(n+1)+}$ be the dominant ionization state of some species. In that case
the most important opacity source (at the wavelengths of interest) is usually provided by
$X^{n+}$. The population of the levels in this ionization state scales with the square of the density.

Figure~\ref{fig_spec_sm_cl} shows the spectral evolution for models Bsm and Bcl
during the photospheric phase.
The spectral differences between our smooth and clumped models Bsm and Bcl are
subtle. The redder colors reported earlier (bottom panels of Fig.~\ref{fig_photometry})
arise from a stronger metal line blanketing in the clumped model, driven by a lower ionization
and higher density in the spectrum formation region.
This supersedes the counter-acting effect of a greater photospheric temperature.
Consequently, the clumped model shows a lower flux shortward of H$\alpha$.
Metal lines like Ba\two\,6142\,\AA\ appear stronger, while the H$\alpha$ emission
flux is reduced. This likely stems from the smaller ejecta volume containing partially-ionized
hydrogen in the clumped model.

It is not surprising that clumping does not strongly affect colors and spectra during the
recombination phase since the photosphere sits at the interface between ionized
and recombined hydrogen (i.e., the recombination front), irrespective of whether the
material is clumped or smooth. The effect of clumping is merely to cause the photosphere
to recede faster to deeper ejecta layers.
Spectroscopically, because of the strong chemical mixing and because lines form over extended
regions (in radius or in velocity), the precise location of the photosphere cannot be
easily deciphered from spectral line widths. In other words, clumping (as implemented
in \cmfgen\ at present) does not lead to spectral differences that can be easily
identified. The lack of clear clumping signatures does not imply clumping is absent,
nor that it should be ignored.

\begin{figure}
  \includegraphics[width=0.98\hsize]{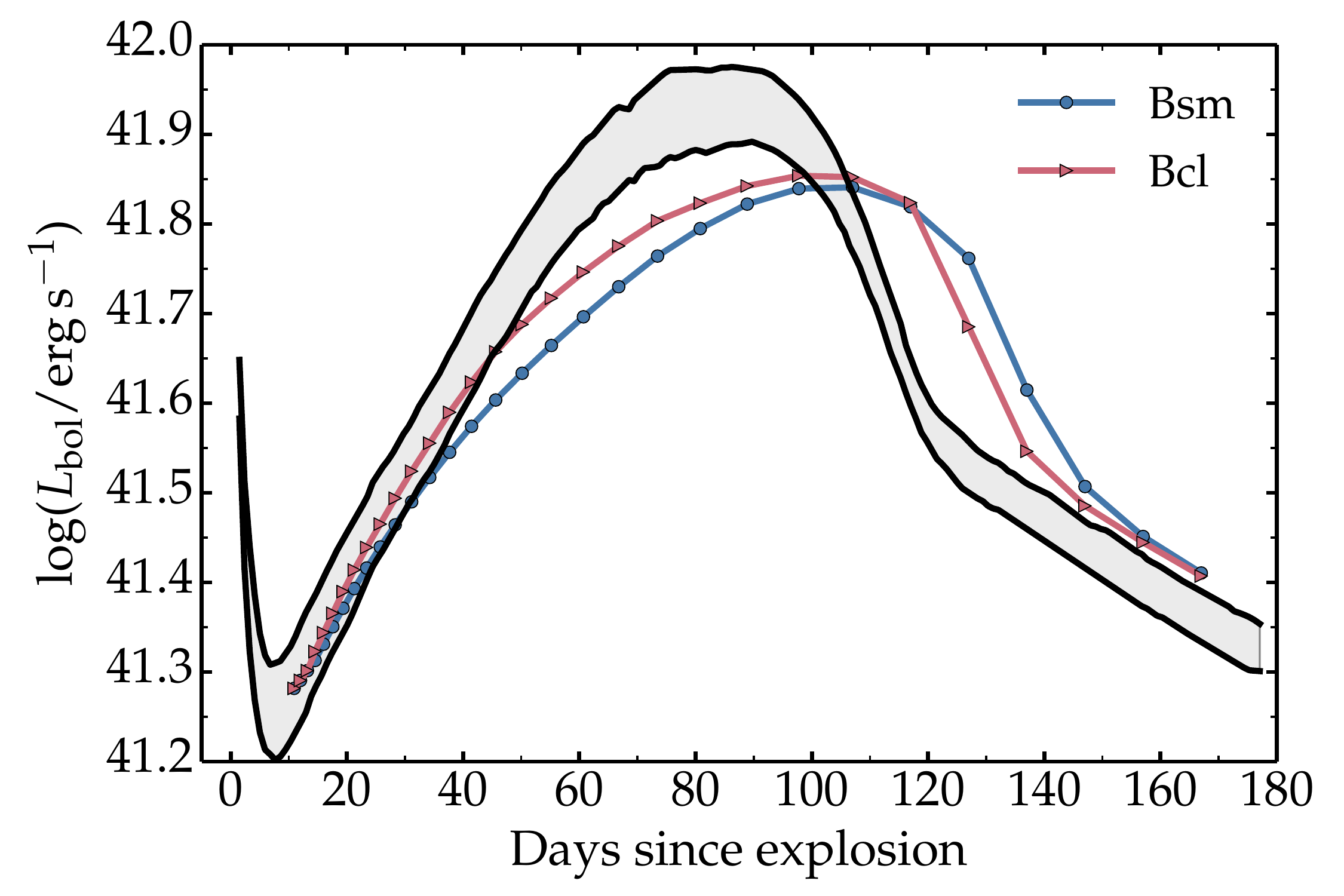}
  \includegraphics[width=0.98\hsize]{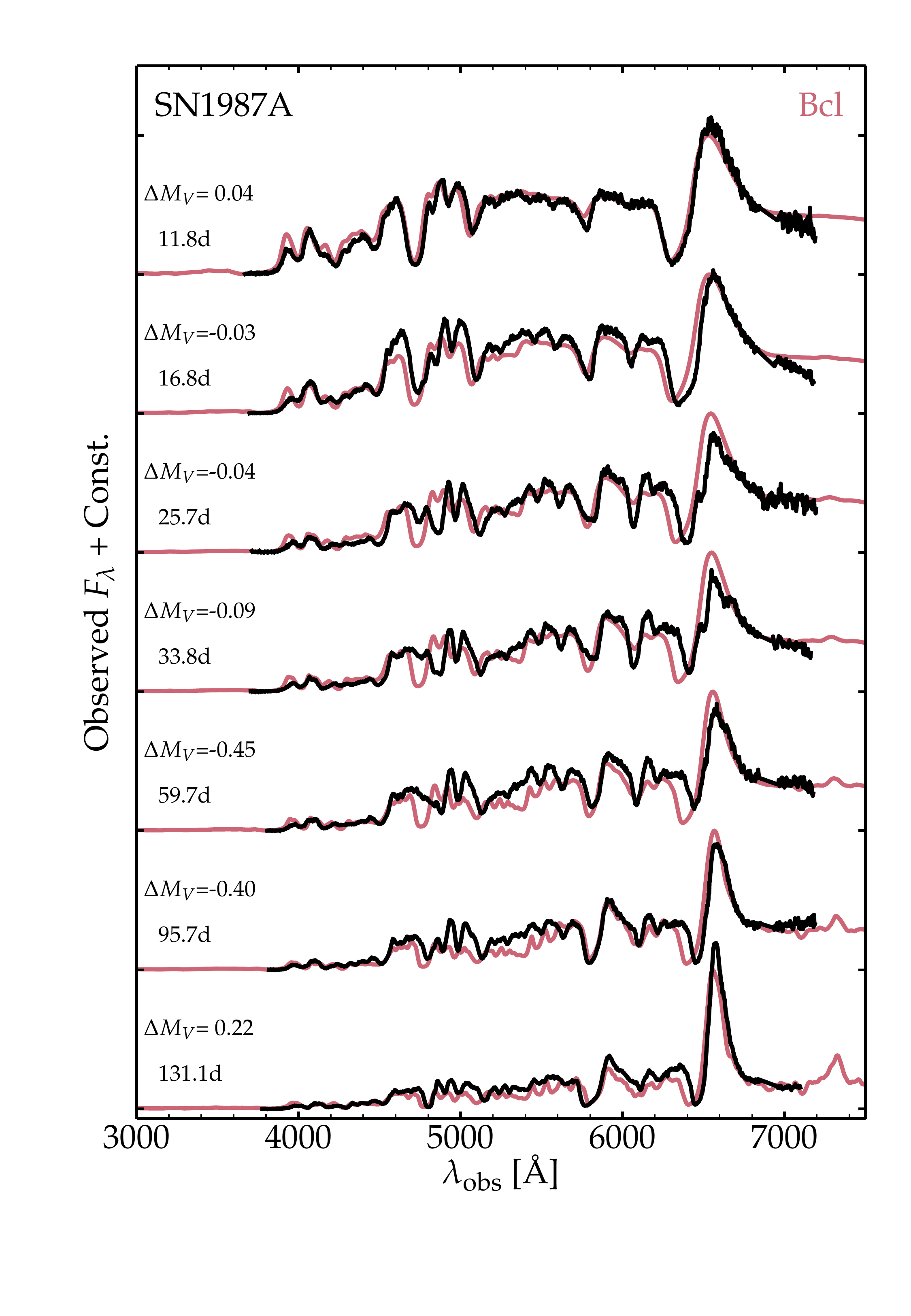}
  \vspace{-1.0cm}
  \caption{Comparison of the bolometric light curve (top; the shaded area is bounded by the
    bolometric light curves inferred by \citet{catchpole_87A_87} and \citet{hamuy_87A_88})
    and multi-epoch spectra
    for SN\,1987A and our BSG explosion model Bcl (redshifted and reddened).
    The label gives the post-explosion epoch and the $V$-band magnitude offset
    between model Bcl and SN\,1987A.
    \label{fig_87A}
  }
\end{figure}

\section{Comparison to observations}
\label{sect_comp_obs}

The BSG model used here is part of a broad investigation on Type II-peculiar SNe
and the diversity of their light curves and spectra (Dessart et al., in prep.).
None of the SN models is specifically designed to match any particular observation.
In other words, we do not iterate  until a given model matches a specific SN. Nonetheless,
it is fruitful to compare theoretical models
to observed BSG-star explosions and in particular SN\,1987A.

SN\,1987A has been modeled extensively in the past, including its light curve properties
(e.g. \citealt{woosley_87a_88}, \citealt{blinnikov_87A_00}; \citealt{utrobin_etal_15})
and its spectral properties (e.g. \citealt{hoeflich_87A_88};
\citealt{eastman_87A_89}; \citealt{schmutz_87A_90}; \citealt{mitchell_56ni_87A_01}).
However, its photometric and spectroscopic properties
have very rarely been modeled simultaneously \citep{UC05,DH10,li_etal_12_nonte}.
In this section, we compare our clumped model Bcl to the observations of SN\,1987A.
We use the bolometric luminosity inferred by \citet{catchpole_87A_87}
and \citet{hamuy_87A_88},
which differ mostly through the uncertain value of reddening.
We use the spectroscopic observations of \citet{phillips_87A_88}. For our spectral
comparisons, we adopt a reddening $E(B-V)=$\,0.15\,mag and an explosion date coincident
with the neutrino detection reported by \citet{hirata_87A_87}.

Figure~\ref{fig_87A} shows the bolometric light curve and spectral evolution
from about 10\,d until the onset of the nebular phase for model Bcl and for SN\,1987A
(for the light curve, both models Bsm and Bcl are shown).
The huge mixing employed (Fig.~\ref{fig_prog_comp}) allows the model Bsm
to continuously rise after 10\,d, as observed in SN\,1987A. With weak mixing,
the rise would be delayed and a kink would be present at about 20\,d.
With no \nifs, the model light curve would instead precipitously drop after 20\,d
(see e.g. \citealt{blinnikov_87A_00}).
In the clumped model, the rise is faster and the time of maximum is earlier,
reducing the offset with SN\,1987A. A greater helium mass fraction or a smaller
radius can also hasten the photosphere recession, but this takes away nothing
from the effect caused by clumping.

Spectroscopically, models Bcl and Bsm are identical at 11.8\,d and reproduce
well SN\,1987A. The subsequent evolution is however perplexing and suggest
that something very unusual is going on. Indeed, the H$\alpha$ line becomes
weak and very narrow on the way to bolometric maximum, H$\beta$ even disappears,
while Ba\two\,6142\,\AA\ and Ba\two\,6496.9\,\AA\
(and probably Ba\two\,4934\,\AA, which may cause the
disappearance of H$\beta$) become surprisingly strong. \citet{UC05} argued that
the strong Ba\two\ lines were a result of time-dependent ionization but this
is unlikely the solution since we treat this process here. In practice, accounting for
time dependence yields a greater ionization than for steady state, which
tends to prevent rather than facilitate the formation of Ba\two\ lines here.
At 35\,d, the dominant Ba ion is Ba$^{2+}$, which explains why Ba\two\
lines are weak.  Increasing the Ba abundance by a factor 5 at 35\,d induces
only a minor change to the synthetic spectrum. The weakness of Ba\two\ lines is therefore
caused mostly by over-ionization, which clumping can affect favorably,
although insufficiently, in model Bcl.
The Ba ionization is also sensitive to the radiation field shortward of 9.3\,eV,
a region in which there is little flux. Additional species and line blanketing might
help reduce the radiation field in this region, and may thus solve the Ba\two\ line
strength problem (as also discussed by \citealt{UC05}).
\cmfgen\ sometimes reproduces the strength of Ba\two\ lines, like in the low-energy
explosion Type II-Plateau SN\,2008bk \citep{lisakov_08bk_17}.

The lower ionization in SN\,2008bk arises from the greater
density in the spectrum formation region (even without clumping), which is a consequence
of the low expansion rate (i.e., the SN ejecta are denser at a given post-explosion time
compared to a standard energy explosion). Interestingly, in SN\,2008bk (and in fact in
all low-energy low-luminosity SNe II-Plateau; \citealt{lisakov_ll2p_18}),
the effect of Ba\two\,6497\,\AA\
on the H$\alpha$ profile is as strong as seen here for SN\,1987A.
Compared to SN\,1987A, a disagreement persists but is reduced when clumping is introduced.
Throughout the evolution, the Na\one\,D is well matched, perhaps because the Na\one\
ionization potential is only 5.14\,eV, and hence the Na ionization state is less sensitive to the EUV
radiation field ($E > 10$\,eV).
As the Na\two\ ionization potential is very large (47.5\,eV), Na\one\ persists over
a wide range of conditions.

In general, to produce a brighter Type II SN display during the photospheric phase,
one increases the explosion energy. But as can be seen in Fig.~\ref{fig_87A},
our model spectra match observations at 10\,d, and overestimate the widths
of Fe\two\,5169\,\AA,  Na\one\,D, or H$\alpha$ later on.
Increased \nifs\ mixing would not help either.
While it would reduce the mismatch in early-time brightness, it would increase the mismatch
in line profile width because of the enhanced non-thermal excitation and ionization in the outer ejecta.
Instead, observations suggest that the SN photosphere is receding faster, a property
that is compatible with clumping. While qualitatively adequate, the clumping we
use in model Bcl is too small to yield a good match to the SN\,1987A spectra
at $30-100$\,d, and to the bolometric light curve. As pointed out earlier, we made
no attempt at reducing the ejecta mass or changing the ejecta composition to yield
a better match. Given all the uncertainties and the simplistic treatment of clumping
in \cmfgen\ at present, our goal is merely to demonstrate that clumping has an effect.

\section{Discussion}
\label{sect_disc}

We have presented non-local thermodynamic equilibrium
time-dependent radiative transfer simulations for the ejecta
resulting from a BSG explosion. The basic ejecta properties
are broadly compatible with SN\,1987A.

The goal of the paper was to describe the impact of ejecta
clumping on SN radiation properties. With our 1-D treatment
of clumping, which leaves unchanged the radial column density,
the rate of recession of the photosphere is increased because of the
greater recombination at the photosphere. The greater material
density also leads to enhanced blanketing and redder optical
colors. The impact on the spectral morphology is subtle, with a
slight reduction of the H$\alpha$ emission and slight increase
in metal line strength. Of all observables, the bolometric luminosity
(or optical flux) is the most influenced, through an increase of
at most $10-15$\,\% at 50\,d, a shorter rise to maximum, and
an earlier transition to the nebular phase. This effect of clumping
is analogous, for example, to what would occur if the ejecta had
a lower mass or a greater helium mass fraction.
The spectral peculiarities of SN\,1987A at 30\,d (abnormally weak H$\alpha$,
absent H$\beta$, strong Ba\two\ lines) cannot be explained by
time dependence, and are not compatible with a stronger \nifs\ mixing.
They are, however, compatible with the effect of clumping, which in our
simulations is probably too small -- the clumping factor at the photosphere
at 30\,d is only 0.65 (over-density by a factor of 1.55 compared to the smooth model
counterpart).

Clumping, as currently implemented in \cmfgen, leads to a higher
recombination rate, causing the faster recession of the photosphere and the
faster release of stored internal energy. This is the main effect captured here.
However, because our clumping is 1-D, it
does not alter the (radial) column density
and thus underestimates the impact that  clumping would have (e.g.
by introducing porosity). Even with clumping, we continue to assume
chemical homogeneity while, for example, the \nifs\ rich material should be under-dense
relative to the H-rich material. A porous medium in 3-D would thus impact
the mean free path of $\gamma$-ray and optical photons differently. Porosity
will exacerbate the effect of clumping described here.
The linear polarization detected at early times in some Type II-Plateau SNe
\citep{leonard_12aw_12} suggests the presence of inhomogeneities in the outer ejecta, thus
further out than adopted here.

We have performed a similar  exploration for the RSG explosion model m15mlt3
of \citet{d13_sn2p}. We computed two such models of Type II-Plateau SNe (Rsm and Rcl)
in an analogous fashion to what is described above for model Bsm and Bcl.
Model Rsm is smooth (identical to m15mlt3) and model Rcl is clumped
(we adopted $a=$\,0.1 and $b=$\,3000\,\kms). We found a very similar behaviour to
Bsm versus Bcl simulations, although here the effect at early times is invisible
due to the large SN brightness. However, at the end of the plateau (corresponding to
the epoch of bolometric maximum in a Type II-peculiar SN), model Rcl is brighter by about 10\,\%
and it also falls off the plateau (i.e., it enters the nebular phase) 10\,d earlier.
Clumping might thus help resolve a problem with (smooth) \cmfgen\ SN II-Plateau models, because these tend to underestimate the brightness at the end of the plateau \citep{d13_sn2p}.
It also suggests that clumping can impact the inferred ejecta mass of Type II-Plateau SNe
since the ejecta mass is in part constrained by the plateau (or the photospheric phase)
duration.

Our smooth BSG explosion model Bsm systematically underestimates the Ba\two\ line
strength observed in SN\,1987A, in spite of our time-dependent non-LTE treatment
\citep{UC05,D08_time}.
Increasing the Ba abundance by a factor of 5 has little impact on the predicted line strengths,
primarily because the ionization of Ba remains too high.
Clumping is a powerful means to reduce the ejecta ionization and enhance the
strength of Ba\two\ lines after 20\,d.

With time dependence, the ejecta ionization is maintained higher, which tends
to inhibit recombination. In the simulations of \citet{D08_time}, assuming
steady state yields an ejecta more recombined, hence with a lower radial optical depth.
This process may be one reason why our \cmfgen\ simulations tend to yield a longer
photospheric phase than in seemingly identical radiation hydrodynamics simulations
with a simplified treatment of the gas \citep{utrobin_etal_15}. This issue
requires further study.

   In this study, the properties for clumping (depth dependence and magnitude)
were largely ad-hoc. Enhancing the magnitude and extent of clumping exacerbates
its impact. In the future, we will need to perform a systematic study to quantify the
impact of clumping for different clumping properties. We will also need to
seek constraints from high-resolution 3D explosion simulations, in which both
large scale and small scale inhomogeneities are resolved.

The effects of clumping discussed here for Type II SNe may be relevant for
other SN ejecta and SN types.
Although not modeled here and therefore speculative,
clumping may help to resolve the
discrepancy between the mass inferred for SN Ibc \citep{drout_11_ibc}
and those expected from single W-R stars. The luminosity boost
caused by clumping in our Bcl and Rcl models may also explain why some
standard SNe Ibc have abnormally large inferred \nifs\ masses \citep{drout_11_ibc}.
At present, while one can explain easily the production
of Type IIb/Ib SNe through the binary channel \citep{yoon_ib_17}, the origin
of Type Ic SNe remains debated. Continued evidence supports the notion
that SNe Ic must come from higher-mass progenitors, including single
W-R stars \citep{maund_snibc_18}.
Clumping, which to this day has never been treated in light curve
calculations, may afford larger ejecta masses that may help resolve
this long standing problem.

\begin{acknowledgements}

We thank Roni Waldman for providing the blue-supergiant progenitor model.

\end{acknowledgements}

\end{document}